\documentclass[useAMS,usenatbib,usegraphicx]{mn2e}

\usepackage{times}

\title[Infrared SED Model of Young Dwarf Galaxies]{
  Infrared Spectral Energy Distribution Model for Extremely 
  Young Galaxies}

\author[T. T. Takeuchi et al.]
  {Tsutomu T.~Takeuchi,$^{1}$\thanks{
    Postdoctoral Fellow of the Japan Society of the Promotion of Science.
    }\thanks{E-mail: takeuchi@optik.mtk.nao.ac.jp.}
   Hiroyuki Hirashita,$^{2}$\thanks{
    Postdoctoral Fellow of the Japan Society of the Promotion of 
    Science for Research Abroad.
    }
   Takako T. Ishii,$^3$
\newauthor
   Leslie K. Hunt,$^4$
   and Andrea Ferrara,$^5$\\
$^1$National Astronomical Observatory of Japan, 2--21--1,
       Osawa, Mitaka, Tokyo, 181--8588, Japan\\
$^2$INAF-Osservatorio Astrofisico di Arcetri, Largo Enrico Fermi, 
       5, 50125, Firenze, Italy\\
$^3$Kwasan Observatory, Kyoto University, Yamashina-ku, Kyoto,
       607--8471, Japan\\
$^4$Istituto di Radioastronomia-Sez. Firenze/CNR, Largo Enrico Fermi, 5, 50125
        Firenze, Italy\\
$^5$SISSA/International School for Advanced Studies, Via Beirut, 4  
       34014 Trieste, Italy\\
}

\date{Released 2002 Xxxxx XX}

\def\sbs{SBS~0335$-$052}
\def\izw{I~Zw~18}

\begin{document}

\label{firstpage}

\maketitle

\begin{abstract}
The small grain sizes produced by Type II supernova (SN II) models
in young, metal-poor galaxies make the appearance of their infrared 
(IR) spectral energy distribution (SED) quite different from 
that of nearby, older galaxies. To study this effect, 
we have developed a model for the evolution of dust content
and the IR SED of low-metallicity, extremely young galaxies 
based on \citet{hirashita02}.
We find that, even in the intense ultraviolet (UV) radiation field of very 
young galaxies, small silicate grains are subject to stochastic heating 
resulting in a broad temperature distribution and substantial MIR 
continuum emission.
Larger carbonaceous grains are in thermal equilibrium at 
$T \simeq 50 \mbox{--} 100$~K, and they also contribute to the MIR.
We present the evolution of SEDs and IR extinction of very young,
low-metallicity galaxies. 
The IR extinction curve is also shown.
In the first few Myrs, the emission peaks at
$\lambda \sim 30\mbox{--}50\,\mu$m; at later times  
dust self-absorption decreases the apparent grain temperatures, shifting 
the bulk of the emission into the submillimetre band.
We successfully apply the model to the IR SED of \sbs,
a low metallicity (1/41 $Z_\odot$) dwarf galaxy with
an unusually strong MIR flux. 
We find the SED, optical properties and extinction of the star forming 
region to be consistent with a very young ($\mbox{age} \simeq 6.5 
\times 10^6\,\mbox{yr}$) and compact ($\mbox{radius} \simeq 20\,\mbox{pc}$) 
starburst.
We also predict the SED of another extremely low-metallicity
galaxy, \izw, for future observational tests.
We estimate the FIR luminosity of \izw\ to be low as $L_{\rm FIR}
\sim 10^{7\mbox{--}7.5}\,L_\odot$, depending on the uncertainty of dust mass. 
Some prospects for future observations are discussed.
\end{abstract}


\section{Introduction}

Galaxy formation is undoubtedly one of the desiderata in modern 
astrophysics, but remains a long-standing unsolved problem.
The quest for `primeval galaxies' is still a strong motivation of deep
surveys by large telescopes and future space missions.
By now, much effort has gone into the search for galaxies at high redshift 
($z$), some of which we believe are experiencing their first episode of 
star formation:
Lyman break objects \citep[e.g.,][]{madau96,steidel96,steidel99};
Lyman $\alpha$ emitters \citep[e.g.,][]{thompson95,hu98,steidel00};
submillimetre galaxies \citep[e.g.,][]{hughes98,eales99,eales00,blain99,
barger00,scott02};
hyper-extremely red objects (HEROs) which are also thought to be strongly
enshrouded by dust \citep[e.g.,][]{totani01}.

However, these objects have already produced significant amount of stars,
and consequently, are chemically enriched by heavy elements.
Galaxies are thought to be extremely metal-poor just after their birth, 
hence they may have had only a small amount of dust. 
Even in Lyman break galaxies however, there is clear evidence that they 
contain non-negligible amount of dust
\citep[e.g.,][]{adelberger00,sawicki01,shapley01}. 
It remains difficult to explore the physics of metal-deficient first 
objects in the early Universe because of our observational limitations.

Among related issues, the extent to which very early star formation has been 
hidden by dust is one of the most interesting themes.
Little doubt remains about the importance of extinction
properties to evaluate high-$z$ star formation \citep[e.g.,][]{steidel99}.
Dust grains absorb stellar light and re-emit it in the far infrared (FIR). 
Indeed, the FIR spectral range represents a unique opportunity to 
study the dust properties and distribution, and several dedicated 
infrared (IR) space or stratospheric missions are planned 
({\sl ASTRO-F}\footnote{http://www.ir.isas.ac.jp/ASTRO-F/index-e.html},
{\sl SIRTF}\footnote{http://sirtf.caltech.edu/},
SOFIA\footnote{http://sofia.arc.nasa.gov/},
{\sl Herschel Space Observatory}\footnote{http://astro.estec.esa.nl/First/}, 
etc.).

Furthermore, how dust forms and evolves in primeval galaxies needs to be
considered in order to understand the chemical and thermodynamical
evolution of a metal-poor interstellar medium (ISM). 
Dust grains drastically accelerate the formation rate of
molecular hydrogen (H$_2$), expected to be the most abundant molecule
in the ISM \citep[e.g.,][]{glover03}.
Hydrogen molecules emit vibrational-rotational lines.
It is the only coolant in a cool (temperature $T \ll 8000$~K) dense core of 
metal-deficient system, since the atomic cooling does not work effectively.
Hence, the cooling rate of a primeval galaxy consequently depends strongly
on the dust content and H$_2$ abundance.

Recently \citet[][hereafter H02]{hirashita02} have developed a model 
for the evolution of dust content in very young (or primeval) galaxies.
Dust is expected to be produced at the final stages of stellar
evolution, and Type II supernovae (SNe II) are the dominant source for
the production of dust grains in young star-forming galaxies
\citep[e.g.,][]{dwek80}.
Dust formation in stellar winds from evolved low-mass stars like
red giant branch stars (RGBs) or asymptotic giant branch stars (AGBs) 
can also contribute considerably, but the cosmic time is not long
enough for such stars to evolve at high-$z$.
On the other hand, dust is also destroyed by SN shocks 
\citep[e.g.,][]{jones96};
however, H02 have shown that the dust destruction is negligible compared to
the formation rate in young ($\sim 10^6 \mbox{--} 10^8$~yr) galaxies.
Then they have modeled the evolution of FIR luminosity and dust temperature 
in such a young starburst on the basis of SNe II grain formation model
of \citet[][TF01]{todini01}. 

While high-$z$ primeval galaxies are still beyond the ability of
present experiments, starbursts in the Local Universe are often assumed to
be good representatives of the star forming activity at high redshifts.
As such, they are cosmologically interesting as key laboratories
for studying the ISM, the transport of supernova processed metals, and
the chemical evolution of galaxies and of the intergalactic medium
\citep[][]{madau01}.
H02 have paid especial attention to the nearby, extremely
low-metallicity galaxy \sbs, since this object may be truly
experiencing its first burst of star formation ($\mbox{age}\la 10^7$ yr; 
\citealt{vanzi00}). 
The models by H02 successfully explain its observed properties such as dust 
mass and temperature, and far infrared (FIR) luminosity.

In this paper, we further investigate the IR properties of very young
galaxies, and construct a simple model of mid- to far-infrared (MIR
and FIR) SEDs of galaxies starting from the H02 models.
As already mentioned above, SNe II are expected to dominate the dust formation
in very young galaxies.
The size of dust grains formed in SNe II cannot be as large as $0.1 \mbox{--} 
1\,\mu$m (TF01), and host galaxies are too young for grains to grow in the 
interstellar space.
TF01 have shown that the size distribution of grains produced by SN II 
is well approximated to be single-sized.
A narrow size distribution favoring small grain sizes makes the appearance 
of IR spectral energy distribution (SED) of young galaxies drastically 
different from that of nearby aged galaxies.
We, for the first time, properly consider the dust size distribution 
peculiar to the very early stage of galaxy evolution, and construct 
a model of the IR SED of very young galaxies.
Then, we compare our models to the observations in \sbs, and continue the 
discussion by H02 on evolution of dust content.

In order to discuss the effects of dust in such galaxies, we must first
discuss the effect of dust sublimation. 
Dust grains are expected to sublime in {\sc H\,ii} regions because of 
the strong radiation from the central source (in this case, OB stars). 
Indeed, the small-sized carbon grains (or PAHs) are deficient in a strong 
ultraviolet (UV) radiation environment \citep[e.g.,][]{boselli98,dale01a}.
However, the radius of the dust cavity caused by the sublimation is much 
smaller than that of the star forming region for classical size grains 
\citep[][and references therein]{inoue02}.
Further, as will be seen later, although our model dust grains are smaller 
than the classical grains, the sublimation process is governed by 
the energy of each injecting photon, because of the small grain cross section.
For small grains, the peak temperature cannot reach the sublimation
temperature.
Hence we can safely neglect the effects of sublimation in our discussion. 

The paper is organized as follows:
We first model the dust content of a young galaxy whose age is less than 
$10^8$ yr in \S\ref{sec:model}.
We show the evolution of the SEDs of young galaxies and their IR
extinction curve in \S\ref{sec:results}.
Then, we discuss some topics related to the infrared properties of young
galaxies in \S\ref{sec:discussion}.
We finally compare the model predictions with the observed quantities of 
the extremely low-metallicity BCD, \sbs, in \S\ref{subsec:sbssed}.
The lowest metallicity star-forming galaxy ever known, \izw\ 
($1/50\,Z_\odot$), is another representative of the star forming dwarfs.
We present a prediction for the SED of \izw\ in \S\ref{subsec:izwsed}.
Based on these investigations, we estimate how many low-metallicity dwarf 
galaxies and blue compact dwarfs will be detected in future observational 
projects in \S\ref{subsec:future}.
The last section (\S~\ref{sec:conclusion}) is devoted to our conclusions.

\section{SED Model for Extremely Young Galaxies}\label{sec:model}

\subsection{Star formation and dust production}

First we briefly describe the star formation and dust production model
proposed by H02.
This model treats the evolution of dust content in a young galaxy whose
age is less than $10^8$~yr. 
Hence dust formation in stellar winds from evolved low-mass stars like RGBs 
or AGBs \citep[e.g.,][]{schmid_burgk80,deguchi80,woodrow82,gail84,gail85,
gail87} is negligible because the time interval is too short for 
stars to evolve to such stages.
We can therefore safely assume that only SN II contributes to 
the dust formation.
For similar reasons, the contribution of Type Ia supernovae to the 
total supernova rate can be also neglected. 
Further, since dust destruction by SNe may be important on a longer timescale 
of $>10^8$~yr \citep[][]{jones96,hiraferrara02}, 
we can also safely neglect this process.
Therefore, the only contributor to the total dust mass ($M_{\rm dust}$) 
in a young ($<10^8$~yr) galaxy is the supply from SNe II.

The rate of SNe II is given by
\begin{eqnarray}
  \gamma (t)=\int_{8~M_\odot}^{\infty}
  \psi (t-\tau_m)\, \phi (m)\, dm\, ,
\end{eqnarray}
where $\psi (t)$ is the star formation rate (SFR) at age $t$ (we define $t=0$ 
at the beginning of the star formation), $\phi (m)$ is the initial mass 
function (IMF), $\tau_m$ is the lifetime of a star whose mass is $m$, 
and it is assumed that only stars with $m>8~M_\odot$ produce SNe II. 
H02 assume a constant SFR, $\psi =\psi_0$, and a Salpeter IMF 
($\psi (m)\propto m^{-2.35}$; the mass range of stars is 0.1\mbox{--}60 
$M_\odot$) to obtain a first estimate\footnote{
We assume a relatively smaller mass cut (60 $M_\odot$) for the stellar IMF,
because TF01 only give dust production rate for stars $\leq 35~M_\odot$. 
Though some assume that the upper mass bound $M_{\rm up}=120\,M_\odot$, 
our framework is insensitive to the upper stellar cut-off, because
the number of such high-mass ($\ga 60\,M_\odot$) stars is
extremely small. 
The contribution of the high-mass stars to UV luminosity is
also much smaller than that of the intermediate ($3\mbox{--}40\,M_\odot$)
mass stars if the stellar luminosity is weighted with the Salpeter IMF. 
In summary, the upper mass limit does not significantly 
affect our conclusions.}.
Then the rate of increase of $M_{\rm dust}$ is written as
\begin{eqnarray}
  \dot{M}_{\rm dust}=m_{\rm dust}\gamma\, ,
\end{eqnarray}
where $m_{\rm dust}$ is the typical dust mass produced in a SN II.
We adopt $m_{\rm dust}=0.4~M_\odot$ (HF02). 
We start the calculation from $M_{\rm dust}=0$ at $t=0$. 
Because dust production rate is proportional to $\gamma$ which is 
proportional to the SFR $\psi_0$, $M_{\rm dust}$
calculated by our model directly scales as $\psi_0$.
The dust mass evolution with
$\psi_0=1~M_\odot~{\rm yr}^{-1}$ is shown in Figure~1 of H02.

\subsection{SED construction}

In this subsection, we present the construction of the SEDs for very young 
galaxies.
As already discussed above, in their early phase, SN II explosions dominate
the supply of dust. 
The galaxies considered here are too young to have evolved low-mass stars 
like RGBs or AGBs.
These evolved low-mass stars are considered to be the source of 
large-size grains with a power-law like size distribution
\citep[e.g.,][]{biermann80,dominik89}.
It should also be noted that they are too young to make small grains grow
larger in their diffuse interstellar medium 
\citep[see, e.g.,][pp.223--224]{whittet92}.

We recall that dust grains are schematically divided into silicate and 
carbonaceous ones; only these two species are considered in this work.
TF01 have shown that the sizes of silicate and carbonaceous grains formed in 
SN II ejecta to be about 10~\AA\ and 300~\AA, respectively.
Then the overall size distribution of dust is well represented by 
a superposition of two delta functions, which correspond to the sizes of 
silicate and carbonaceous grains.
TF01 estimated the uncertainty of the grain sizes to be a factor of $\sim 2$ 
within a reasonable range of the relevant thermodynamical parameters.
Other sources of uncertainty and their estimates are discussed in TF01.
Here we should keep in mind the lack of large-size grains of $\sim 0.1 
\mbox{--} 1\,\mu$m in very young galaxies, while a substantial contribution of 
such large `classical' grains is inferred from the observation of 
the Galactic diffuse IR emission or the IR SEDs of nearby galaxies.
The discrete and small grain sizes make the appearance of the SED of 
young galaxies drastically different from that of aged normal galaxies, 
especially in the IR.
We next focus on the most important radiation process of small dust grains.

\begin{figure*}
\centering\includegraphics[angle=90,width=\textwidth]{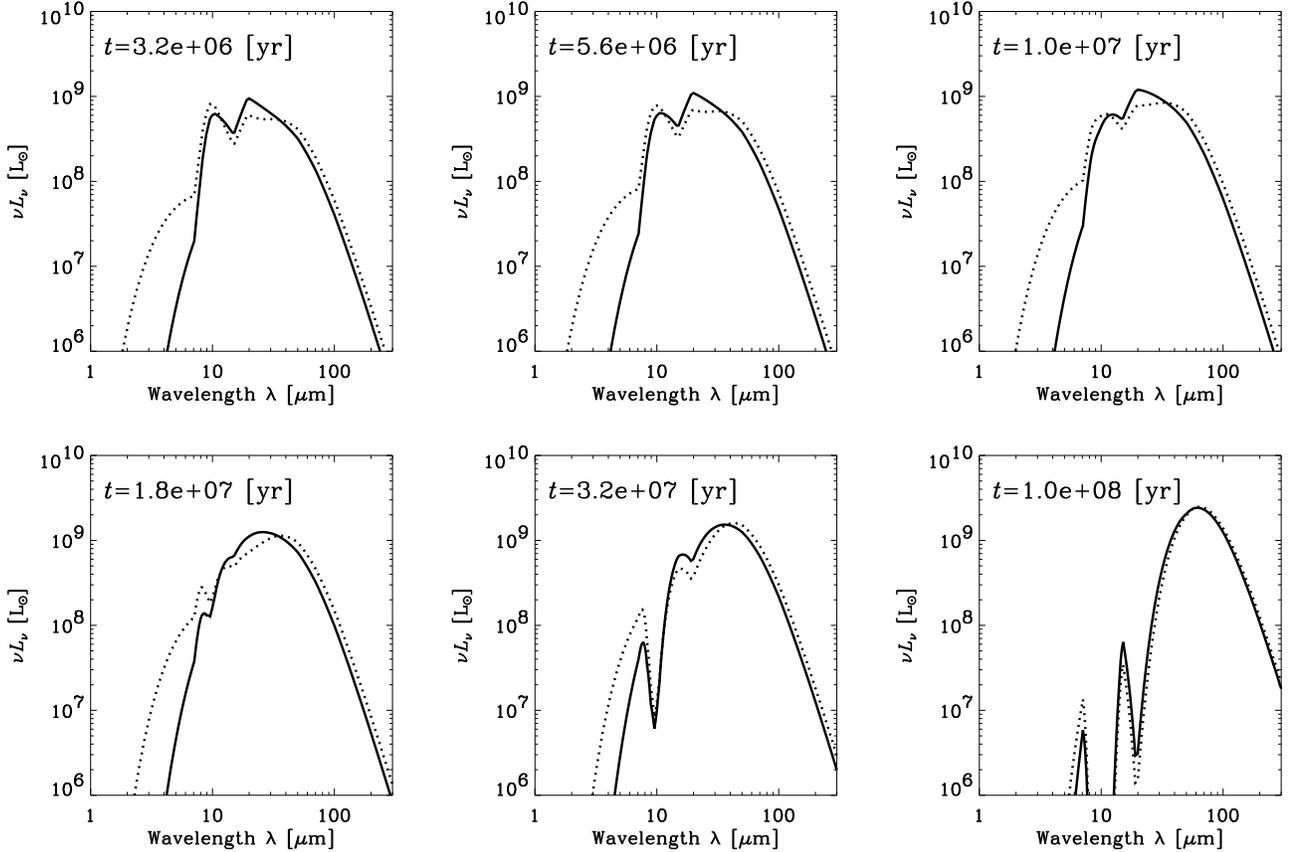}
\caption{Evolution of IR SED of a very young galaxy. 
 The size of the star forming region is  $r_{\rm SF}=30$~pc.
 Solid and dotted lines represent the SED history for a galaxy with
 the size of silicate grain $a_{\rm sil} = 10$~\AA\ and 6 \AA,
 respectively. 
 For Figures~\ref{fig:irsed_30pc} and \ref{fig:irsed_100pc}, 
 constant star formation rate of $\mbox{SFR}=1\,M_\odot$ is assumed.
}\label{fig:irsed_30pc}
\end{figure*}

\begin{figure*}
\centering\includegraphics[angle=90,width=\textwidth]{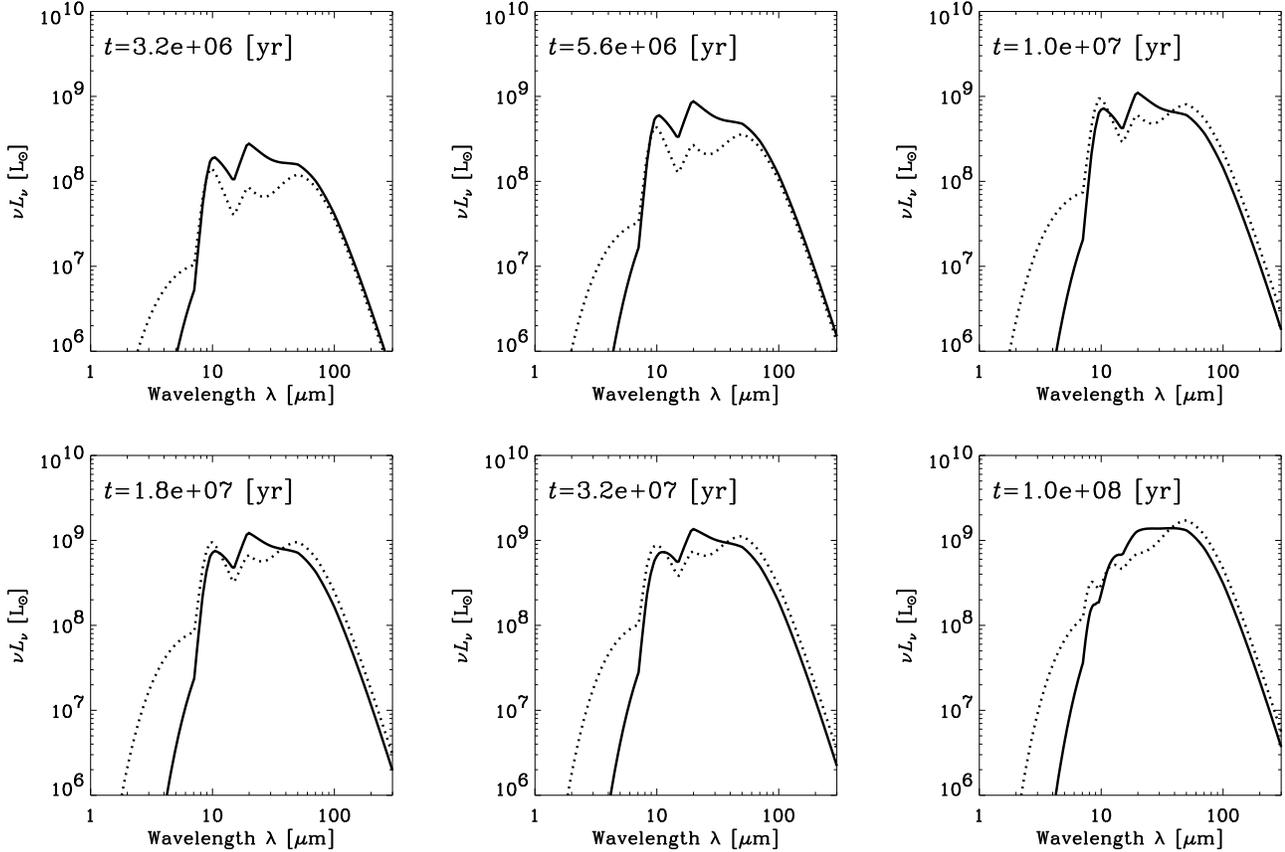}
\caption{Same as Figure~\ref{fig:irsed_30pc} but the size of the star 
 forming region is  $r_{\rm SF}=100$~pc.
}
\label{fig:irsed_100pc}
\end{figure*}

\subsubsection{Stochastic heating of very small grains}

The dust grains originating from SNe II cannot be as large as those formed 
in the atmosphere of evolved AGB stars or grown in the diffuse interstellar 
space (\citealt{kozasa87,kozasa89}; TF01).
TF01 clearly showed that the sizes of the silicate grains and of the
carbonaceous grains are much smaller than that of classical large grains.

Here we note that TF01 used the classical nucleation theory for the 
formulation of mass accretion onto grains \citep{feder66}.
These grain sizes might be even smaller if we use a more recent formation
theory of dust grains from metal-rich gas medium \citep[e.g.,][and 
references therein]{ford97,tanaka02a}.
This revision may introduce many subtle and perplexing effects in the 
dust formation process, and further investigations are awaited 
\citep[cf.][]{tanaka02b}.
The effect of dust size will be addressed later in the context of 
the origin of near-infrared (NIR) continuum radiation.

It is well accepted that very small grains are `stochastically heated',
that is they cannot establish thermal equilibrium with the ambient radiation 
field (\citealt{purcell76,aannestad79,
draine85,guhathakurta89,siebenmorgen92,li01,draine01} (DL01)):
the heat capacity of very small grains is too small to maintain its
temperature until the next photon impinges.
Hence, the equilibrium temperature of a grain cannot be well 
defined because its `temperature' varies violently with time, i.e., 
it becomes very hot each time a UV photon is absorbed by the grain, 
and subsequently cools rapidly to a very low temperature.
Consequently, the continuum shape from very small grains becomes much
wider than that of a modified blackbody, and the continuum has a strong
extension to the NIR.

We adopt this picture and use it to calculate the dust emission in 
the NIR--MIR.
Now we formulate the stochastic heating of very small dust grains.
The heat capacity of a grain has been extensively studied and modeled by 
previous papers (DL01; \citealt[][and reference therein]{li01}).
We apply the multidimensional Debye model for the heat capacity of 
a grain, $C(T)$, as introduced by DL01.\footnote{As the authors have already 
pointed out in their subsequent paper, there is a typographic error in
their Equation~(10) in DL01.}
The specific heat of silicate grains is represented by 
\begin{eqnarray}
  C_{\rm sil}(T) = (N_{\rm atom} -2)k\left[ 
    2 f_2' \left( \frac{T}{500 \, \mbox{K}}\right) +
    f_3'\left( \frac{T}{1500 \, \mbox{K}}\right)\right],
\end{eqnarray}
and that of carbonaceous grains is
\begin{eqnarray}
  C_{\rm C}(T) = (N_{\rm C} -2)k\left[ 
    f_2' \left( \frac{T}{863 \, \mbox{K}}\right) +
    2 f_2'\left( \frac{T}{2504 \, \mbox{K}}\right)\right],
\end{eqnarray}
where 
\begin{eqnarray}
  f_n (x) &\equiv& n \int_0^1 \frac{y^n\,dy}{\exp (y/x) -1} \, , \\
  f_n' (x) &\equiv& \frac{d}{dx}f_n(x)\, ,
\end{eqnarray}
and $k$ is the Boltzmann's constant (the subscripts `sil' and `C' represent 
the silicate and carbonaceous grains, respectively).
Namely, as for silicate grains, two-thirds of the vibrational 
modes are distributed according to a Debye model with $n=2$ and Debye 
temperature $\Theta = 500$~K, and one-third of the modes are described by 
a Debye model with $n=3$ and $\Theta = 1500$~K.
For carbonaceous grains $N_{\rm C} -2$ out-of-plane modes and
$2(N_{\rm C}-2)$ in-plane modes are set.
This model excellently reproduces $C(T)$ of each grain species 
(see Figure~1 of DL01).
We adopt $\rho_{\rm C} = 2.26\, [\mbox{g\,cm}^{-3}]$ \citep{draine84} 
and $\rho_{\rm sil} = 3.5\, [\mbox{g\,cm}^{-3}]$ \citep{li01}, 
i.e., $N_{\rm C}=1.14 \times 10^{23}\, [\mbox{cm}^{-3}]$ and 
$N_{\rm atom}=8.5 \times 10^{22}\, [\mbox{cm}^{-3}]$, 
respectively ($N_{\rm atom}$ is the number density of atoms).
Here we identify bulk graphite grain with our carbonaceous grain.

The (effective) equilibrium temperature, $T_{\rm eq}$ is computed by equating
grain heating by photon absorption and cooling by photon emission
\citep[see Equations~(6.1) and (6.2) of][]{draine84}:
\begin{eqnarray}\label{eq:balance}
  \int_0^\infty u_\lambda c Q(a, \lambda) d\lambda =&& \hspace{-6mm}
    60\sigma \left( \frac{hc}{\pi k}\right)^4 \times \nonumber \\
    &&\hspace{-6mm}\int_0^\infty \frac{Q(a,\lambda) \lambda^{-5}}{
      \exp\left[hc/\left(\lambda kT_{\rm eq}\right)\right]-1}\,d\lambda \,,
\end{eqnarray} 
where $u_\lambda$ is the interstellar radiation field energy density, 
$a$ is a grain size, and $h$, $\sigma$ and $c$ are Planck's constant, 
Stefan--Boltzmann constant, and the speed of light, respectively. 
We adopted the values presented in \citet{draine84} for the
absorption efficiencies, $Q_{\rm abs} (\lambda)$.
In practice, the above equation can be approximated by setting $Q$ at UV
(on the lhs of Equation~\ref{eq:balance}) to be unity and $Q$ at FIR
(on the rhs) to behave like $Q_{\rm FIR} \propto \lambda^{-2}$.
Then we obtain
\begin{eqnarray}
  T_{\rm eq} \simeq \left(\frac{hc}{\pi k}\right)\left\{ \frac{945 u}{960\pi 
    (2\pi Aa) hc} \right\}^{1/6}
\end{eqnarray}
and 
\begin{eqnarray}
  u \equiv \int_0^\infty u_\lambda d\lambda \, , \; 
  Q_{\rm FIR} \equiv \frac{2\pi Aa}{\lambda^2} \,.
\end{eqnarray}
Here we adopt $A_{\rm sil}=1.34\times 10^{-3}\,[\mbox{cm}]$ for silicate 
grains \citep[][]{drapatz77} and 
$A_{\rm C} = 3.20\times 10^{-3}\,[\mbox{cm}]$ for carbonaceous grains
(scaled to fit $Q_{\rm IR}(\lambda)$ of \citealt{draine84}).

The actual dust temperature $T$ has a very wide distribution around 
$T_{\rm eq}$.
As we will see later, while carbonaceous grains can be in equilibrium with
the ambient radiation field, silicate grains cannot establish the equilibrium, 
even in the intense radiation field (high photon density) of a young galaxy 
(\S\ref{sec:results} and \S\ref{sec:discussion}).

\subsubsection{Emission}\label{subsubsec:emission}

Following \citet{draine85},
with this $C(T)$, we calculated the temperature distribution of dust as
a function of size $a$ with Monte Carlo simulations for incident UV photons.

We calculated $u_\lambda$ from the history of OB star luminosity of H02
and the size of the star forming region $r_{\rm SF}$. 
We use the OB star luminosity calculated in H02 as the total luminosity of
stars with $m>3~M_\odot$.
We start from $r_{\rm SF} = 30$~pc as the size of a star-forming region in
young galaxies (H02), but other sizes are also examined later.

We approximated the spectrum of the UV field by the formula of 
\citet{draine96} for simplicity and scale it with luminosity by factor 
$\chi (t)$ as
\begin{eqnarray}\label{eq:uv_field}
  u_\lambda = &&\hspace{-6mm} \left[ -\frac{25}{6} 
    \left(\frac{10^{5}\times \lambda}{1\,\mbox{cm}}\right)^2 + 
    \frac{25}{2} \left(\frac{10^{5}\times \lambda}{1\,\mbox{cm}}\right)
    - \frac{13}{3} \right] \nonumber \\
    &&\hspace{-6mm}\times 10^{-9} \chi(t) \; 
    [\mbox{erg\,cm}^{-3}\mbox{cm}^{-1}] \,.
\end{eqnarray}
The UV radiation field strength of the solar neighborhood,
$u^{\rm NBH}$, is
$u^{\rm NBH} = 4.0 \times 10^{-14}\,[\mbox{erg\,cm}^{-3}]$
\citep{draine96}.
The intensity factor $\chi (t)$ is determined as a function of the galaxy 
age from the OB luminosity history and $r_{\rm SF}$ (Equation~\ref{eq:chi})
\footnote{We note that, for a high-$z$ galaxy, we should include CMB 
in the radiation field.
Small grains spend most of their time with their lowest temperature, which
is determined by CMB.
This slightly affect the submillimetre SED of high-$z$ galaxies.
An extensive discussion on the heating of dust by CMB is given in 
\citet{ferrara99a}.}.

For the geometry, we assume that the OB stars are concentrated 
in the centre of the system, and dust is assumed to be 
distributed as a shell, so as to express the fact that dust absorbs 
the UV photons very effectively in star forming regions \citep{hunt01}.
Moreover, as seen later, such a ``simplest'' geometry is successful
enough to explain the SED of \sbs.
We regard the radius of the sphere, $r$ as the size of the spherical 
star forming region, $r_{\rm SF}$.
Then, the flux is calculated by $J(t)=L_{\rm OB}(t)/(4\pi r_{\rm SF}^2)$.
Then the UV energy density incident on dust grains, $u$, is 
\begin{eqnarray}\label{eq:chi}
  u=\frac{J(t)}{c}=\chi(t) u^{\rm NBH}\,.
\end{eqnarray}
The spectral UV energy density $u_\lambda$ is calculated from the assumed 
spectrum of the UV radiation field (Equation~\ref{eq:uv_field}).

Such a shell-like geometry of dust distribution may result from the drift 
induced by the radiation pressure. 
\citet{gail79} show that grains in H\,{\sc ii} regions can be
accelerated nearly to $v_{\rm d}\sim 10\,\mbox{km\,s}^{-1}$.
The crossing time $t_{\rm cross}$ can be estimated as 
\begin{eqnarray}
  t_{\rm cross} \sim 3 \times 10^6 \left(\frac{r_{\rm SF}}{30~{\rm pc}}\right)
    \left(\frac{v_{\rm d}}{10~{\rm km~s}^{-1}}\right)^{-1} \;[\mbox{yr}]\,. 
\end{eqnarray}
Therefore, the dust grains can be swept into a shell surrounding 
the star-forming region within a timescale of $10^6\,\mbox{yr}$. 
Stellar winds can also sweep dust grains 
\citep[see][and references therein for the creation of
central dust cavity]{inoue02}.
We will go back to this issue in line with the application of our model to 
the local low-metallicity star forming dwarf galaxies.

The rate at which a grain absorbs a photon with energy $E \sim E+dE$
is expressed as
\begin{eqnarray}\label{eq:prob}
  \frac{d^2 p}{dEdt} = Q_{\rm abs} (a, \lambda)\pi a^2 u_\lambda 
    \frac{\lambda}{h^2c} \,.
\end{eqnarray}
The heating is represented as follows:
\begin{eqnarray}\label{eq:heating}
  \frac{hc}{\lambda} = \frac{4\pi}{3}\int_{T_0}^{T} C(T')dT' \, ,
\end{eqnarray}
where $T$ is the peak temperature achieved by a grain hit by a 
photon with energy $hc/\lambda$, and $T_0$ is the grain temperature 
just before absorption.   
If more photons are absorbed by a grain simultaneously, the lhs will 
become the total energy of the photons.
On the other hand, the grain cools through radiation as 
\begin{eqnarray}\label{eq:cooling}
  \frac{d T}{dt} = -\frac{3\pi}{aC(T)} 
    \int_0^\infty Q_{\rm abs}(a,\lambda)\, B_\lambda(T) d\lambda \,.
\end{eqnarray}
{}From Equations~(\ref{eq:prob}), (\ref{eq:heating}), and (\ref{eq:cooling}),
we obtain the sample path of a grain temperature as a function of time.
A sufficiently large set of Monte Carlo realizations yields the dust 
temperature distribution \citep{draine85}.
Though there could be faster algorithms \citep{guhathakurta89}, 
we adopt this direct method in order to examine the cooling 
behavior and timescale of very small dust grains in an extremely intense 
radiation field of a young galaxy.

The total mass of each grain component is given by TF01.
The mass ratio we adopt here is $M_{\rm sil}:M_{\rm C} = 0.56:0.44$
(cf.\ H02).
With this value and material density of each species ($\rho_{\rm
sil}=3.50\,\mbox{g}\,\mbox{cm}^{-3}$ and $\rho_{\rm
C}=2.26\,\mbox{g}\,\mbox{cm}^{-3}$), we obtain total numbers of
grains, $N_i$
\begin{eqnarray}\label{eq:dust_number}
  N_i = \frac{3M_{\rm dust}f_i}{4\pi a_i^3\rho_i} \, , 
\end{eqnarray}
where subscript $i$ denotes the species of dust, sil or C, and $f_i$ 
is the mass fraction of dust of species $i$.

Monochromatic luminosity of a grain with radius $a$ and species $i$, 
$L_{\rm IR}^{{\rm grain}, i}(\lambda)$, is, 
then, obtained by
\begin{eqnarray}
  L_{\rm IR}^{{\rm grain}, i}(\lambda) = 
    4\pi a_i^2 \pi\int Q_{\rm abs}^i (\lambda) B_\lambda(T)
    \frac{dP_i(\chi,a_i)}{dT}\,dT \,,
\end{eqnarray}
where $dP_i(\chi,a_i)/dT$ is the temperature probability density function.
In the present case, $a_i$ is fixed for each dust component to
a specific value in the following range:
$a_{\rm sil} = 6\mbox{--}10\,\mbox{\AA} = 
6\mbox{--}10\times10^{-8}\,\mbox{cm}$ and 
$a_{\rm C} = 200\mbox{--}300\,\mbox{\AA} = 
2\mbox{--}3\times10^{-6}\,\mbox{cm}$,
i.e., the dust size distribution function is well 
approximated by a Dirac's delta function\footnote{
The results will not be affected by the detailed choice of the functional form:
e.g., Gaussian with small standard deviation, instead of delta function, 
changes the results little.
The discrete nature of the size distribution is fundamental here, 
i.e., slowly varying functions such as a power-law yield
significantly different results in contrast.
}
for each species, $i$,
\begin{eqnarray}
  \frac{dn_i}{da}(a)=n_i\delta(a-a_i)\,.
\end{eqnarray}

Total IR luminosity from species $i$, $L_{\rm IR}^i$, is
\begin{eqnarray}
  L_{\rm IR}^i \hspace{-3mm}&=&\hspace{-3mm} 
    N_i \int L_{\rm IR}^{{\rm grain},i} (\lambda) d\lambda \nonumber \\
    \hspace{-3mm}&=&\hspace{-3mm}4 \pi a_i^2 N_i \pi 
      \int \int Q_{\rm abs}^i(\lambda)B_\lambda(T)
      \frac{dP_i(\chi,a_i)}{dT}\,d\lambda dT \,,
\end{eqnarray}
and we have imposed the constraint
$L_{\rm IR} = \sum_i L_{\rm IR}^i \leq L_{\rm OB}(t)$;
if $L_{\rm IR}$ exceed $L_{\rm OB}$ we set $L_{\rm IR}=L_{\rm OB}$.
Final continuum IR emission from a young galaxy is calculated by 
superposing each of the continuum from silicate and carbonaceous grains
as a function of starburst age, $t$,
\begin{eqnarray}
  L_{{\rm IR},\lambda}(t) \hspace{-3mm}&=&\hspace{-3mm}
    \sum_i L_{\rm IR}^i (\lambda) \nonumber \\
    \hspace{-3mm}&=&\hspace{-3mm} \sum_i 4 \pi a_i^2 N_i \pi 
    \int Q_{\rm abs}^i(\lambda)B_\lambda(T)\frac{dP_i(\chi,a_i)}{dT}\,dT \,,
\end{eqnarray}
where in the rhs starburst age is incorporated with $N_i$ and 
$dP_i(\chi,a_i)/dT$ through $M_{\rm dust}(t)$ and $\chi(t)$.

\begin{figure*}
\centering\includegraphics[width=8cm]{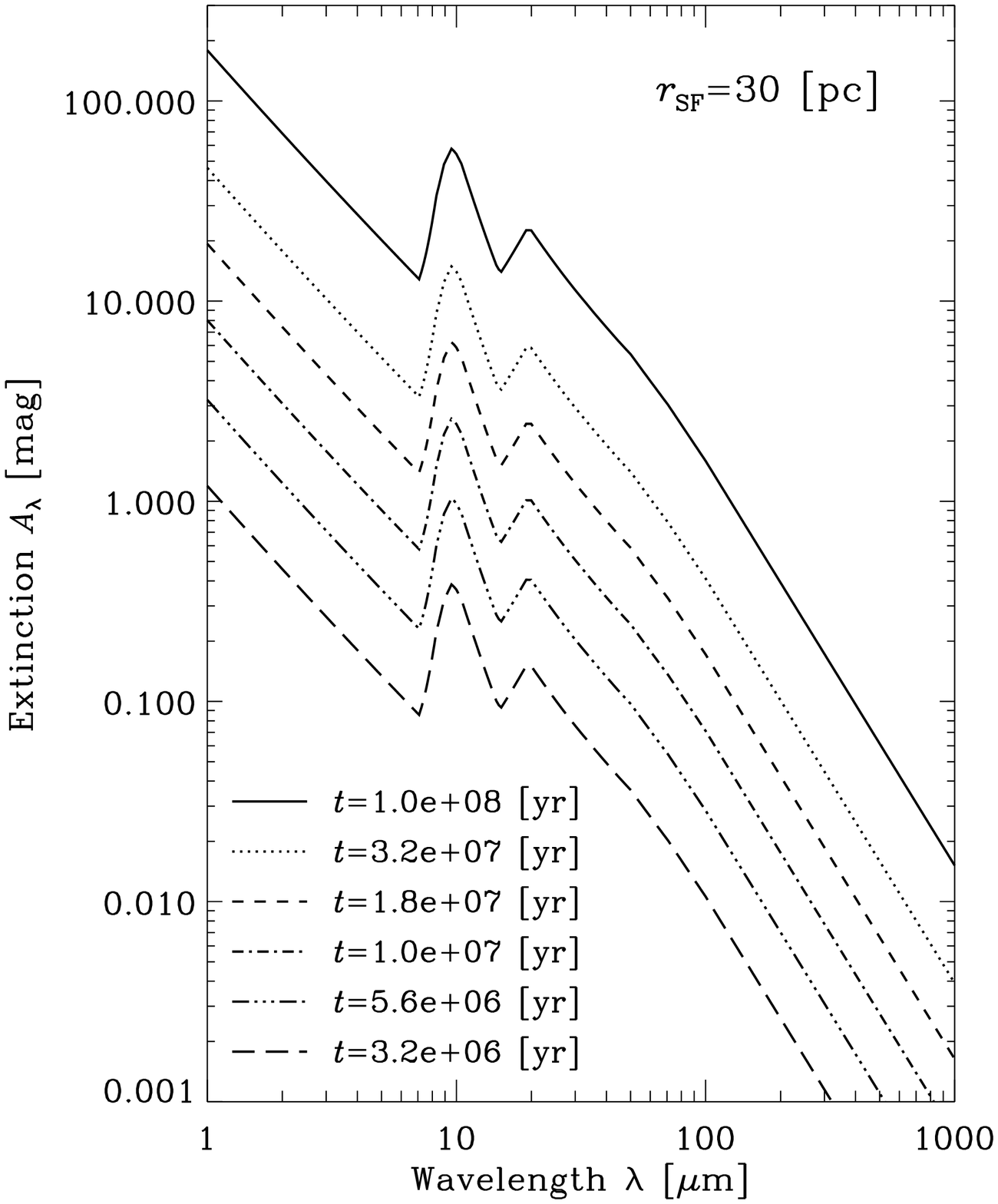}
\centering\includegraphics[width=8cm]{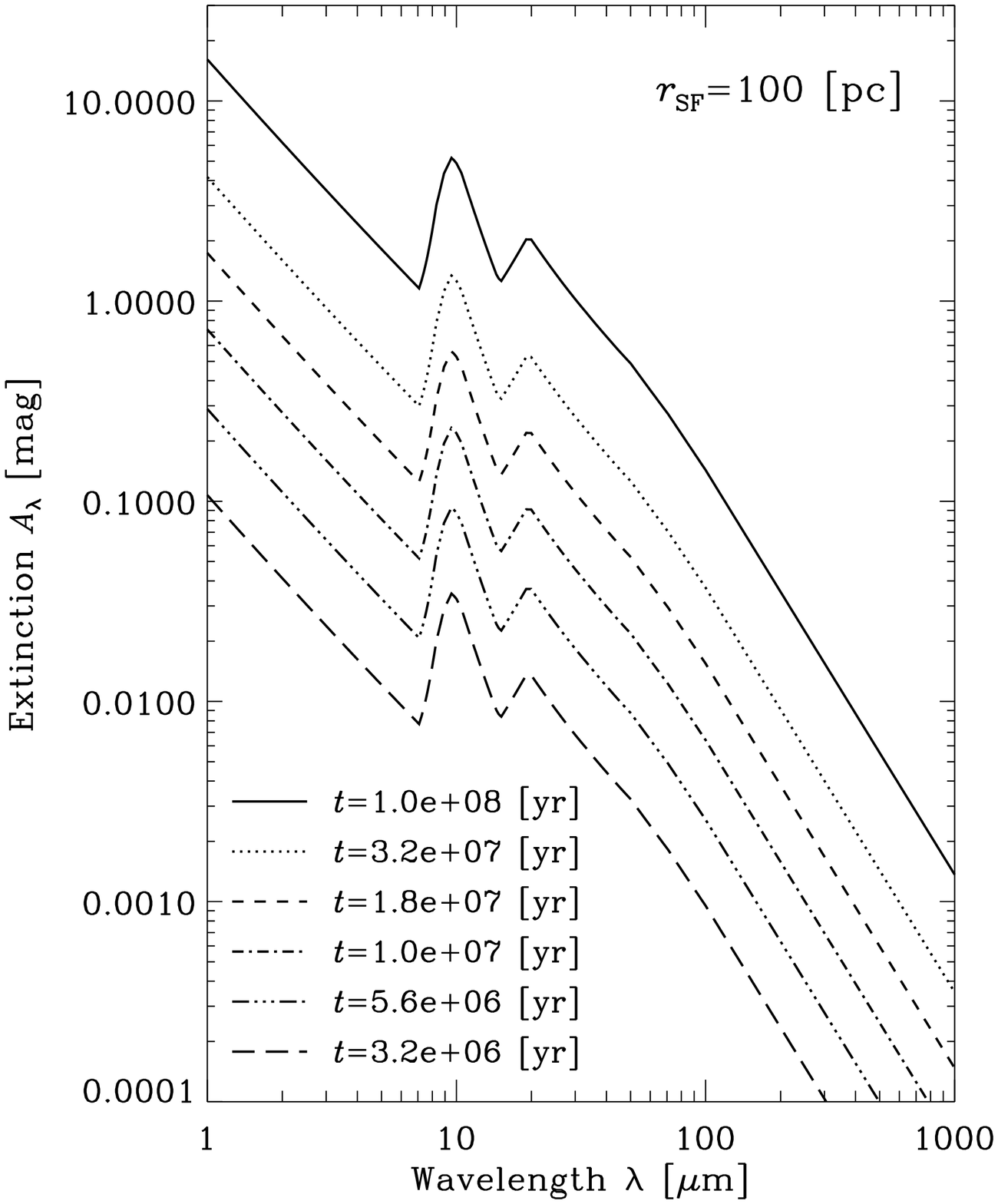}
\caption{Left panel: Evolution of IR extinction curve of a very young galaxy. 
 Size of the star forming region $r_{\rm SF}=30$~pc.
 Right panel: Same as the left panel but $r_{\rm SF}=100$~pc
 (note that the ordinate is an order of magnitude different from that in the
 left panel).
}\label{fig:irextinction}
\end{figure*}

\subsubsection{Extinction}

If the dust opacity is very large, then self-absorption occurs, and 
even MIR radiation from dust is absorbed by the dust itself.
In this work we did not try to solve the detailed radiative transfer but
rather treat the extinction as given by a one-zone screen model. 
The resulting extinction curve is different from that of \citet{draine84} 
because of the different dust size distribution.
In this case dust opacity and extinction are expressed as 
\begin{eqnarray}\label{eq:tau}
  \tau_{\rm dust}(\lambda) \hspace{-3mm}&\simeq&\hspace{-3mm}
    \sum_i \pi a_i^2 Q_{{\rm abs},i}(\lambda)\,n_i r_{\rm SF} \,, \\
  I(\lambda) \hspace{-3mm}&=&\hspace{-3mm} 
    I_0(\lambda) e^{-\tau_{\rm dust}(\lambda)} \,.
\end{eqnarray}
The absorbed light is re-emitted at longer wavelengths, mainly in the 
submillimetre, and consequently, the SED is deformed by the self-absorption.
The final SED is obtained via this self-absorption--re-emission process
with an obvious constraint of the energy emitted in the IR, $L_{\rm IR}$, 
which must be conserved.

The screen geometry might seem an oversimplification, but many
observations support screen-like dust configuration in intense
starbursts \citep[e.g.,][]{meurer95}. 
Moreover, as discussed in \S\ref{subsubsec:emission}, dust can be swept 
into a shell surrounding the star-forming region.
The possible clumpiness of dust may also increase the dust opacity
$\tau_\lambda$ of the galaxy  \citep[for a discussion see][]{ferrara99b}, 
as well as it may allow the UV photons to leak out of the galaxy. 
Although such an anisotropy is difficult to model, it is remarkable that 
the simple geometry we adopted can reproduce a number of the observed 
properties of low metallicity star forming dwarfs as we will see later.
This aspect will be further considered in \S\ref{subsubsec:extinction}.

\section{Results}\label{sec:results}

\subsection{Infrared SED evolution}

We present the results of the evolution of IR SEDs for a very young
galaxy with the age of $10^6\mbox{--}10^8$~yr in 
Figures~\ref{fig:irsed_30pc} and \ref{fig:irsed_100pc}.
To illustrate the results, and facilitate comparison with H02, we
concentrate on the value $\mbox{SFR}=1\,M_\odot\,\mbox{yr}^{-1}$.
We set the silicate dust grain size to $a_{\rm sil}=10$~\AA\ and 
$6\,$\AA\ (solid and dotted lines, respectively), because grain
size can be changed by a factor of about two depending on the conditions of
grain formation.
The size of carbonaceous grains can also be altered for the same reason,
but it affects the SED very little because the grains are in a thermal
equilibrium. 
Thus, we only show the results for $a_{\rm C} = 300\,$\AA\ in 
Figures~\ref{fig:irsed_30pc} and \ref{fig:irsed_100pc}.
In the present model, the radiation field is typically $10^{4\mbox{--}5}$ 
times stronger than in the solar neighborhood.

In the case of smaller silicate dust size, continuum radiation produced 
by stochastically heated grain extends toward NIR wavelengths.
This effect is quite sensitive to the grain size of $a \la 10\,\mbox{\AA}$,
in our case that of silicate grains.
In contrast, the so-called `very small grains (VSGs)' in the 
local normal (viz.\ solar metal abundance) star forming galaxies are 
generally thought to consist of carbonaceous grains or their fragments.
In the early, optically thin phase of a young galaxy, these small silicate 
grains produce a prominent band feature around $9.7\,\mu$m in emission.
It is worth mentioning that even in the extremely intense radiation field 
of $\chi = \mbox{a few} \times 10^4\mbox{--}10^5$, such a small grain 
cannot establish thermal equilibrium with the ambient radiation, because
the cooling time of a grain (typically $t_{\rm cool} \sim10^{-2}$~s 
at $T\sim 500$~K) is much shorter than the mean interval of photon 
injection onto the grain ($\Delta t \sim 10\mbox{--}10^3\,$s).

\citet{plante02} suggested  that the absence of unidentified infrared
band (UIB) features in the intense UV environment of \sbs\ implies that 
stochastically heated small grains are unimportant in such galaxies.
However, as we have explained, small silicate grains considered here 
may be uncorrelated with UIBs.
Small silicate grain is naturally expected in SNe (TF01), so we should 
properly take into account 
the effect of stochastic heating of small grains in starburst systems.
Indeed, until the extinction becomes significant, their contribution is 
dominant in N--MIR wavelengths in our results.

We also stress that the dust temperature of carbonaceous grains
is very high, $T>100$~K in the early phase.
Because of the lack of classical size ($\sim 0.1\mbox{--}1\,\mu$m) large 
grains, submillimetre continuum radiation is weak, and the peaks
are located around the MIR during the first a few tens of Myrs.
Such a high dust temperature is also predicted for giant forming ellipticals
by \citep{totani02}.

\begin{figure*}
\centering\includegraphics[width=8.0cm]{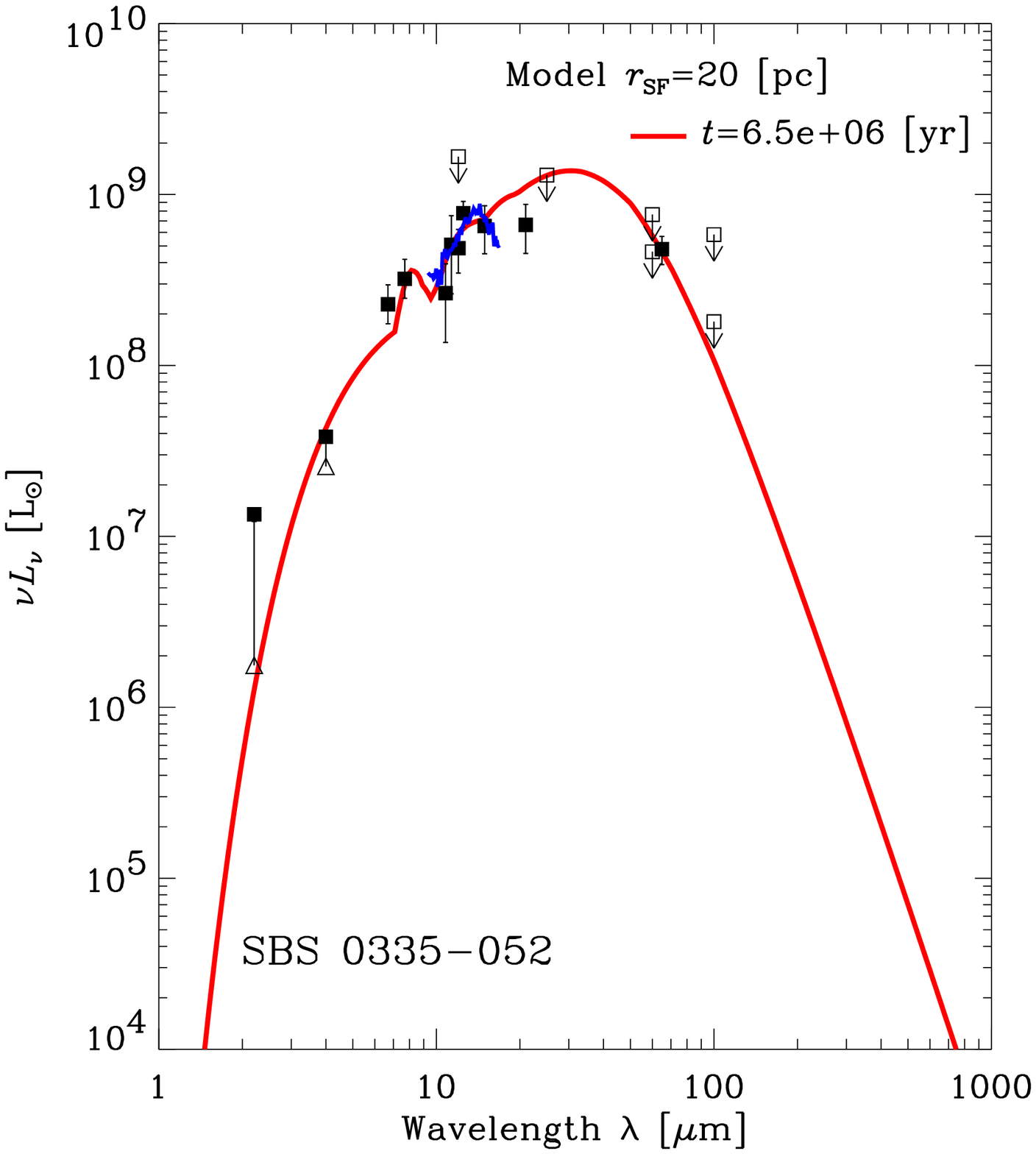}
\centering\includegraphics[width=8.0cm]{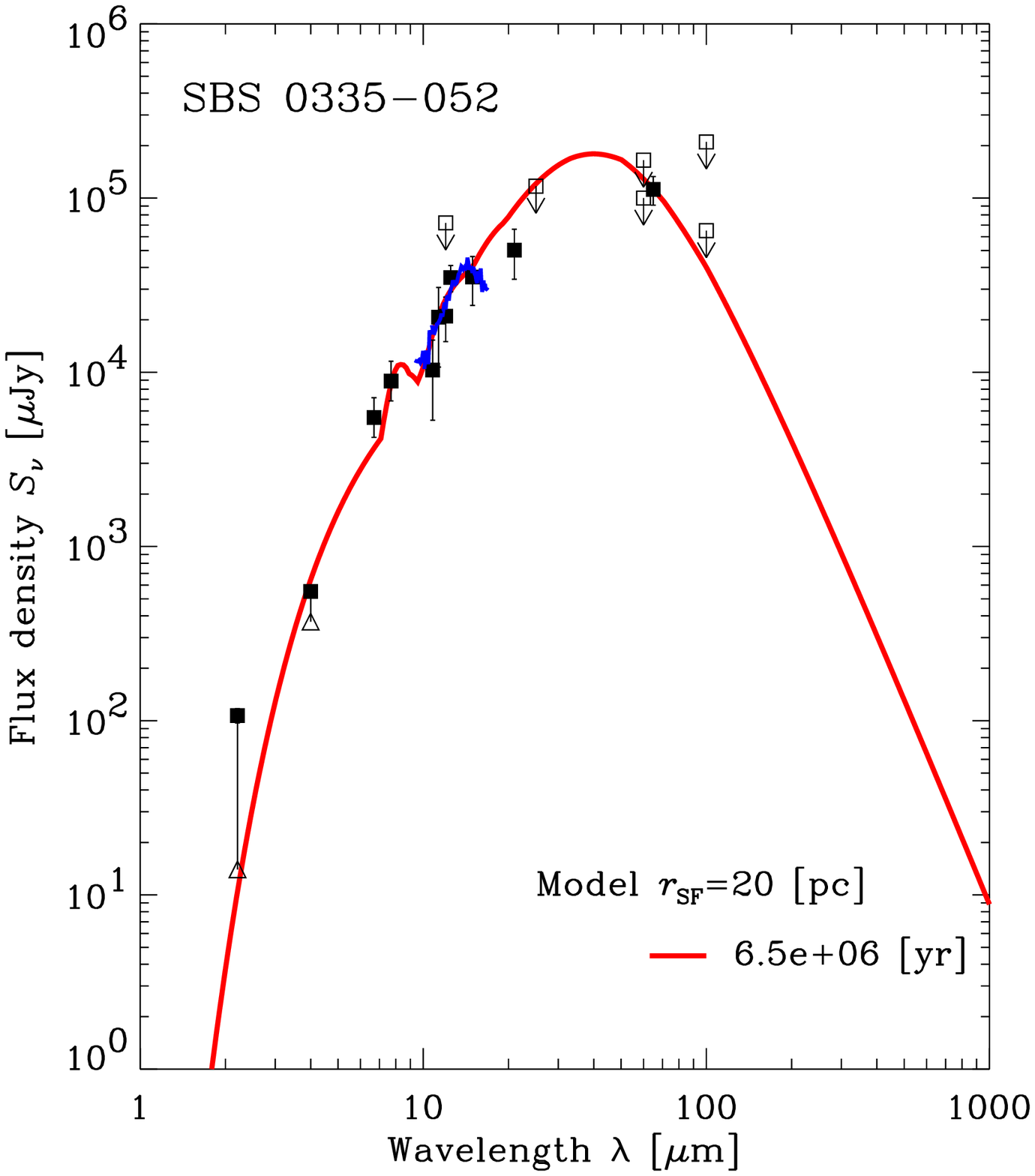}
  \caption{The model SED of an extremely young galaxy calculated by our
  model and the observed SED of \sbs.
  Red curve represents the model SED.
  Filled squares are the observed values, while open squares
  are the upper limits.
  Open triangles in the near infrared are the net dust contributions
  to the total flux (see main text).
  Blue curve is the ISOCAM spectrum obtained by \citet{thuan99}.
  {\sl Left panel}: All photometric values are converted to the monochromatic 
  luminosity $\nu L_\nu\,[L_\odot]$.
  {\sl Right panel}: Same as the left panel but expressed by observed flux 
  densities $S_\nu\,[\mu\mbox{Jy}]$.
}\label{fig:sbs}
\end{figure*}

\subsection{Opacity evolution}

After the early phase characterized by the radiation from hot dust
($\mbox{age} \la 10^7$~yr), 
opacity of dust becomes large, and consequently self absorption of dust gets 
quickly severe in the N--MIR regime.
Finally, the observed dust temperature decreases to 50~K or less.
In particular, the $9.7\,\mu$m silicate band causes a prominent absorption
feature even for a young age $t \sim 10^7$~yr.
The absorbed light is then re-emitted in the FIR-submm wavelengths,
which results in a cooler effective temperature.
Consequently, the temperature history is consistent with that of H02
which was derived simply from energy considerations.

We show the evolution of the IR extinction curves for the cases 
$r_{\rm SF}=30$~pc and 100~pc in Figure~\ref{fig:irextinction}.
We should note that, since dust is supplied only from SNe II 
(i.e., the dust size distribution does not vary)
within the timescale considered here, shape of the extinction curve
does not change in time.
Hence, strictly speaking, Figure~\ref{fig:irextinction} represents
the evolution of dust opacity, rather than the evolution of extinction curve. 

By our definition, $A_\lambda$ is independent of the grain size $a_i$,
since, from Equations~(\ref{eq:dust_number}) and (\ref{eq:tau}),
$A_\lambda$ is obtained as
\begin{eqnarray}
 A_\lambda = 1.086\tau_{\rm dust}(\lambda) 
  \propto \left(\frac{M_{\rm dust}f_i}{\rho_i r_{\rm SF}^2 } \right)
  \frac{Q_{{\rm abs},i}}{a_i}\,,
\end{eqnarray}
and $Q_{{\rm abs},i}$ scales as $Q_{{\rm abs},i} \propto a_i$.
Hence, results are the same for $a_i=6$~\AA\ and 10~\AA.

Here we should keep in mind that the large opacity makes radiation pressure 
effective \citep{ferrara91}, and may cause the expansion of the 
star-forming complex and eventually, the onset of galactic wind at some
stage in the galaxy evolution.
Then the opacity will become smaller than shown in 
Figure~\ref{fig:irextinction}.
In order to take it into account, we should construct a consistent framework
which can treat dynamical and chemical evolution as well as radiation.
We regard dynamical modeling as the next stage and do not 
go further in this work.
Some comprehensive discussions on galactic winds in the similar condition 
can be found in, e.g., \citet{plante02}.
More general properties of the wind in forming dwarf galaxies are discussed 
in, e.g., \citet{ferrara00}.

\section{Discussion}\label{sec:discussion}

\subsection{A starbursting dwarf \sbs}\label{subsec:sbssed}

\subsubsection{Observed SED}

Here we adopt our model to the local BCD, \sbs\ with its unusual IR SED 
and strong flux at MIR in spite of its low metallicity
($1/41\,Z_\odot$).
Since \citet{hunt01} estimated the SFR of this galaxy to be about
$1.7\,\mbox{M}_\odot\mbox{yr}^{-1}$ from $\mbox{Br}\alpha$ line
intensity, we scale the OB star luminosity of H02 with this value.
The result is shown in Figure~\ref{fig:sbs}. 
The photometric data presented in filled and open squares 
in Figure~\ref{fig:sbs} are mainly taken from Table~1 of 
\citet{dale01b}.
The points at $10.8$ and $21\,\mu$m obtained by Gemini/OSCIR
and the data (and more strict upper limits) at 60, 65, and $100\,\mu$m 
obtained by ISOPHOT are from Table~1 of \citet{plante02}.
The ISOCAM spectrum (blue curve) is taken from \citet{thuan99}.
In the NIR, we should correct the contribution of stellar light and 
nebular continuum.
By assuming the stellar light fractions to be $\sim 37$~\% 
($2\,\mu$m) and $\sim 6$~\% ($4\,\mu$m) and applying the nebular emission 
coefficients of \citet{joy88}, 
\citet{hunt01} obtained the net dust contribution of $\sim 13$~\% and 
$\sim 67$~\% at $2\,\mu$m and $4\,\mu$m, respectively.
We show these values by open triangles in Figure~\ref{fig:sbs}.
In the left panel, the data are converted to the monochromatic luminosity 
$\nu L_\nu\,[L_\odot]$, while in the right panel we show in the unit of 
the flux density in Figure~\ref{fig:sbs}.

{}From the NIR spectrum, the age of the stellar population is estimated to 
be $\la 5$~Myr \citep{vanzi00}.
In order to reproduce an extreme NIR--FIR SED of \sbs\ within the strict 
age limit, stellar energy radiated in the optical wavelengths 
must be effectively converted to FIR radiation.
It suggests a very large dust opacity, requiring a very compact SF region 
($r_{\rm SF} = 20$~pc).

The small size of the SF region is also consistent with the observed high
dust temperature $\sim 80$~K.
\sbs\ has a strong N--MIR continuum without UIB or PAH features 
\citep{thuan99}.
In our framework, this is consistent with a small silicate grain size 
$a_{\rm sil}=6$~\AA.
Further, the MIR peak and lack of a prominent submillimetre component strongly 
support the absence of classical-size large grains.
The size of the carbonaceous grains is suggested to be 
$a_{\rm C}=200\,\mbox{\AA}$ in order to reproduce the ISOPHOT observation of 
\citet{plante02}.
\citet{plante02} also reached to a similar conclusion for the
lack of classical size grains from a different radiative transfer model.

A significant absorption feature at $\lambda \sim 10\,\mu$m suggests 
a large dust opacity.
It is a natural consequence of the large IR luminosity of this galaxy, and
is also consistent  with visual extinction of $A_V =12\mbox{--}20$~mag 
for the embedded star forming region \citep{thuan99,hunt01}.
This aspect will be discussed in \S\ref{subsubsec:extinction}.

As we mentioned in \S\ref{subsubsec:emission}, the configuration of dust 
may be shell-like within a timescale of $\sim 10^6$\,yr.
The estimated age of the starburst of \sbs\ is comparable to the crossing 
timescale of the system.

With these physical values and under our assumptions, we have succeeded in 
reproducing the overall SED of \sbs\ with the starburst age $t \sim 6.5$~Myr.
Thus, all the peculiar properties of \sbs\ are consistent with those of 
a genuine young galaxy in the Local Universe.
This success strongly supports the validity of our model, despite 
its simplicity.

\begin{figure}
\centering
\includegraphics[width=9cm]{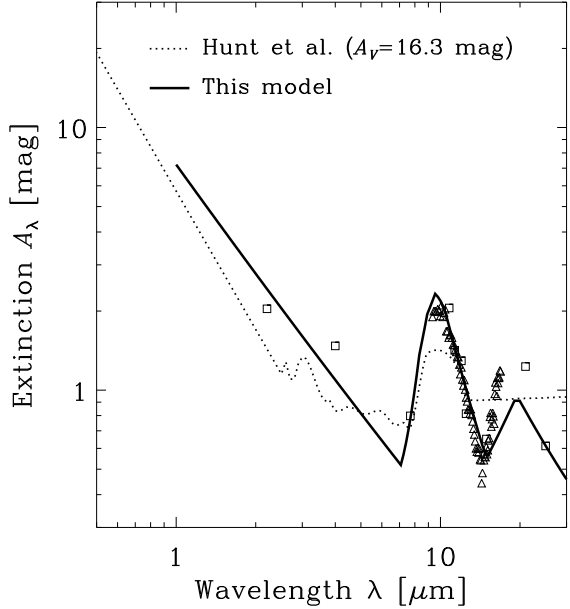}
\caption{
The IR extinction curve obtained from our model, with
age $t \sim 6.5$~Myr (solid line).
The \citet{lutz99} extinction curve, scaled to $A_V\,=\,16.3$~mag
\citep{hunt01} is also shown (dashed line). 
Open symbols represent the extinction deduced semi-empirically from 
the `bare' model SED and the observed data:
squares are the semi-empirical extinction curve derived from 
the photometry, and triangles are from the ISOCAM spectrum.
}\label{fig:extinction}
\end{figure}

\subsubsection{Extinction curve}\label{subsubsec:extinction}

\citet{hunt01} have shown that the extinction properties of \sbs\
can be well described by the extinction curve of \citet{lutz99}, 
advocated for the extinction of the Galactic Centre (GC), assuming
a simplifying assumption of screen-like dust geometry.
In their evaluation of extinction, Hunt et al.\ assumed a superposition of 
two modified blackbodies for the continuum radiation, only the cooler of 
which was extinguished by the screen.
\citet{thuan99} and \citet{dale01b} also assumed modified
blackbody for the unextincted continuum to explore the properties of 
the IR radiation of this galaxy.

However, as we have already seen in the previous section, we cannot expect
equilibrium between the UV radiation field and very small 
dust grains, which are thought to be the main source of the continuum 
radiation in the N--MIR.
This is also pointed out by \citet{dale01b}.
Bearing this in mind, here we reconsider the IR extinction properties. 

The IR extinction curve predicted by our model is presented in 
Figure~\ref{fig:extinction}.
This curve is the one of a galaxy with the same parameters used in 
Figure~\ref{fig:sbs}.

\begin{figure*}
\centering\includegraphics[width=8cm]{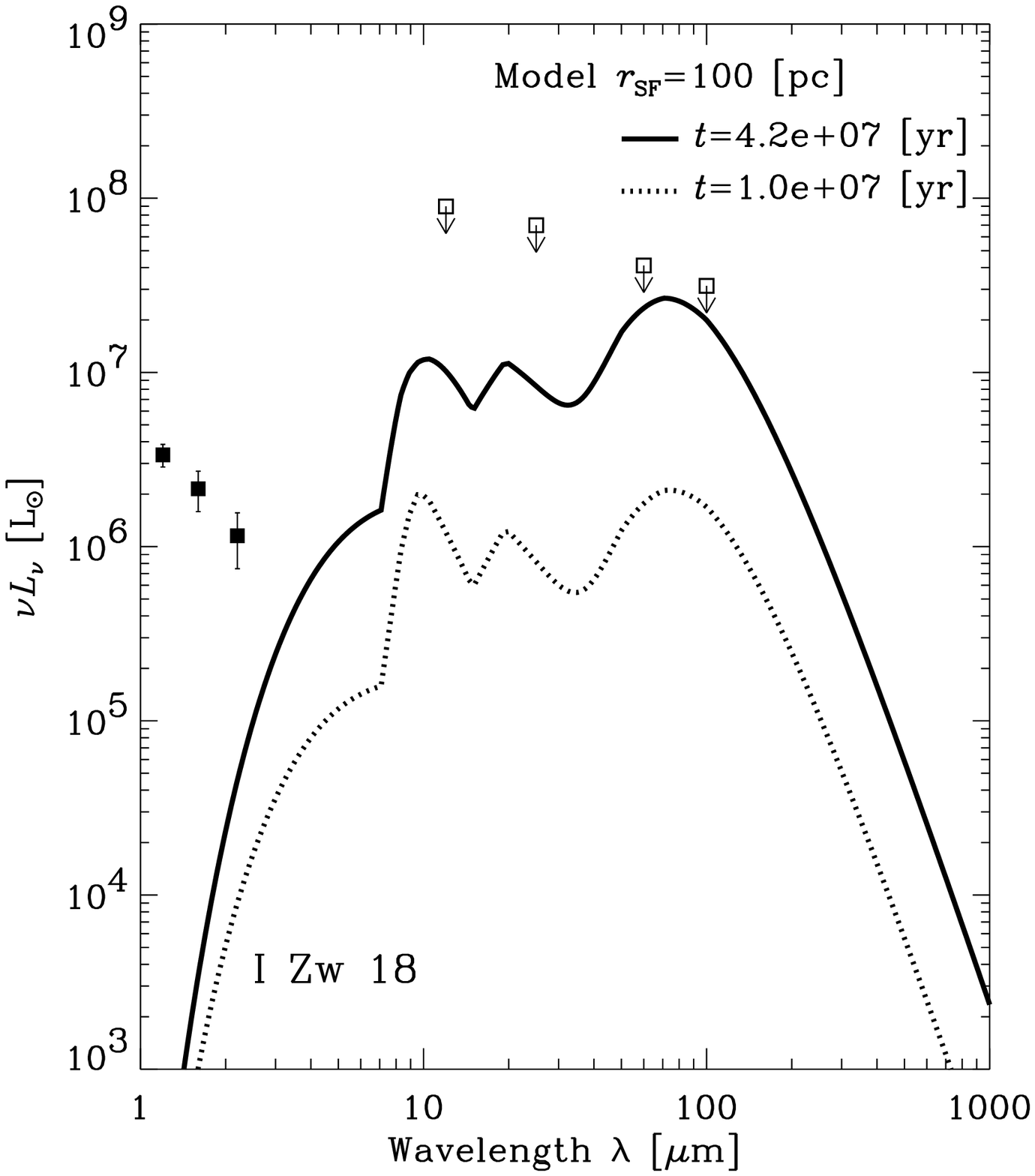}
\centering\includegraphics[width=8cm]{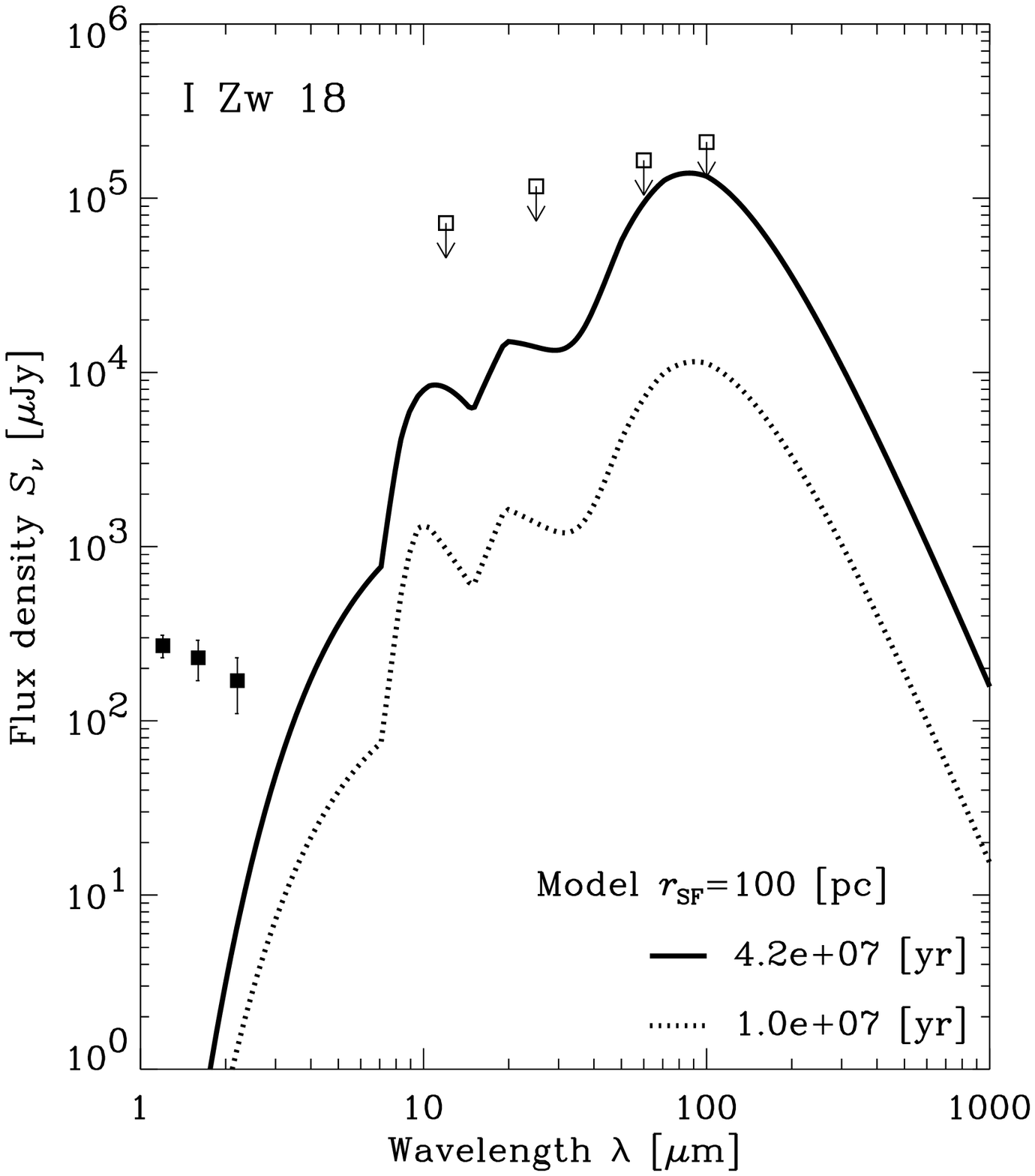}
\caption{The predicted SED for \izw.
  The NIR observed data are taken from \citet{hunt03}.
  Upper limits are calculated from the limits of{\sl IRAS}.
  {\sl Left panel}: The same as Figure~\ref{fig:sbs}, the photometric values 
  are converted to the monochromatic luminosity $\nu L_\nu \, [L_\odot]$.
  {\sl Right panel}: Same as Figure~\ref{fig:izw}, but expressed by observed
  flux densities $S_\nu\, [\mu\mbox{Jy}]$.
}\label{fig:izw}
\end{figure*}

First, it is roughly consistent with the \citet{lutz99} extinction curve, 
scaled by a visual extinction $A_V \simeq 16.3$~mag \citep{hunt01}.
Indeed, the GC curve is quantitatively well reproduced by our model with 
$\mbox{SFR}=1.7\,\mbox{M}_\odot\,\mbox{yr}^{-1}$, 
the star forming region size $r_{\rm SF} \sim 20$~pc, and the above starburst 
age.
Since our model provides the extinction curve only at wavelengths 
$\lambda > 1\,\mu$m, if we extrapolate the model extinction of $K$-band to 
$V$-band by the ratio tabulated in \citet{rieke85},
we obtain the visual extinction strength $A_V \sim 18$~mag.
It is also equal to the value calculated by the relation $A_K\,=\,0.11A_V$
\citep{cardelli89}.
This value is in good agreement with the estimate of Hunt et al., although 
lower than the value $A_V = 30\,$mag by \citet{plante02}.
This may be caused by the difference of the radiation models used for 
calculation.
Here we should keep in mind that the $A_V$ measured at the optically visible
region of \sbs\ is small \citep[$A_V \simeq 0.55$~mag; e.g.,][]{izotov97}, 
typical of dwarf irregulars.
So, the peculiar characteristics of \sbs\ may be governed by the heavily 
embedded star-forming region.

Though the model extinction curve is roughly consistent with the GC
one in the NIR regime, our model predicts a flatter slope than the GC one
in the extrapolation to the optical.
Our extinction curve in the NIR regime has a slope similar to 
\citet{cardelli89} ($\propto \lambda^{-1.6}$).
\citet{hunt01} found that the extrapolation of Lutz's GC curve to the optical
by the slope of $\propto \lambda^{-1.75}$, close to that of \citet{seaton79}
(see the extension of the dashed line to the optical in 
Figure~\ref{fig:extinction}).
The flatter slope of our extinction curve is a natural consequence of the fact
that the continuum radiation is not produced from the dust in equilibrium 
with ambient UV field.
The shape of the non-equilibrium continuum is much broader than that of
modified blackbody radiation.
The intensity of our continuum radiation in the N--MIR is stronger than that
of the modified blackbody if we fix the intensity at about $20\,\mu$m,
hence, we need a stronger extinction (flatter slope) at N--MIR wavelength 
range.
The stronger extinction at the $9.7\,\mu$m broad silicate feature is also
explained in a similar way.
Such behaviour is required to reproduce the absorption feature in the
observed SED of \sbs.

In order to examine these issues in more detail, we estimated the 
extinction with our `bare' model SED, calculated without taking into 
account the self-absorption, rather than using the modified blackbody 
continuum.
The deduced extinction is represented by open squares and triangles 
in Figure~\ref{fig:extinction}. 
We call the newly derived extinction the `semi-empirical extinction curve'.
Unfortunately, since the NIR fluxes from dust emission contain a large
uncertainty depending on the nebular and stellar contribution, we cannot
strongly constrain the shape of the curve, but the NIR part of our model 
extinction curve shows rough agreement with the semi-empirical estimates.
Around the $9.7\,\mu$m silicate feature, the model extinction curve shows 
very good agreement with the semi-empirical one.
This result vindicates the performance of our model, as well as the 
simplifying assumption of a screen-like dust geometry.

However, around the $18\,\mu$m silicate feature, our model significantly 
underestimates the extinction.
Correspondingly, our model SED overproduces the flux around $20\,\mu$m,
compared with the observation of Plante \& Sauvage (Figure~\ref{fig:sbs}).
Many IR galaxies show the strong absorption feature at $18\,\mu$m 
\citep[e.g.,][]{dudley97}.
The discrepancy stems from the shape of the extinction curve around 
$20\,\mu$m.
We cannot reproduce such a feature within our present framework
based on the dust emissivity of \citet{draine84}.
This riddle will be investigated in future work.

Except the above discrepancy, our model extinction curve 
naturally explains most important properties of the GC curve of \citet{lutz99}.
The success of our model in reproducing the extinction curve strongly 
supports the composition of dust supplied by SNe II as
predicted by TF01.

\subsection{Prediction for \izw}\label{subsec:izwsed}

\izw\ has long been a record holder as the lowest metallicity
star-forming galaxy in the Local Universe ($Z=1/50\,Z_\odot$), as well as 
a textbook example of a dwarf starburst.
Hence, in the context of our model, it is an intriguing object to discuss 
as another representative of extremely low-metallicity galaxies.
\citet{ostlin00} estimated its distance to be 12.6~Mpc, and we use this
value for our discussion.
Whether \izw\ is really a newly formed system or not has been a matter 
of debate.
The discovery of AGB stars \citep{ostlin00} may suggest an underlying 
old population, but its contribution is almost certainly not dominant
\citep{hunt03}.
For this reason, we therefore regard \izw\ as a forming galaxy for the first
approximation.

Recently, from HST WFPC2 narrowband imaging, \citet{cannon02} derived 
various important physical properties of this galaxy.
They showed that the O8 ionizing equivalent number of this galaxy is 
2886 for the total galaxy.
It corresponds to a total OB star luminosity, 
$L_{\rm OB}\simeq 4.9\times 10^8\,L_\odot$.
{}From the H$\alpha$ image, we estimate the total size of the SF regions 
of $r_{\rm SF}\sim 50\mbox{--}100$~pc.
The SFR of \izw\ is $\sim 0.04\,M_\odot\,\mbox{yr}^{-1}$, and 
for some patches, $A_V \sim 0.5$~mag \citep{cannon02}.
We use these values to constrain the model prediction.

{}From the measured extinction, \izw\ may also contain a significant 
amount of dust.
\citet{cannon02} estimated the total dust mass 
$M_{\rm dust} \simeq 2\mbox{--}5\times 10^3 \, M_\odot$.
This value is consistent with the dust-to-gas ratio expected
from the metallicity \citep{lisenfeld98,cannon02}.
The dust mass of our model reaches this value at the starburst age
$t \sim 10^{7\mbox{--}7.5}$~yr, an age consistent with the observationally 
suggested major SF age in \izw\ \citep[e.g.,][]{martin96,hunt03}.
Here, we note that a smaller size of the SF region makes $A_V$ exceed 
0.5~mag and is not permitted.

We show the resulting SED in Figure~\ref{fig:izw}. 
The photometric data in the NIR are taken from \citet{hunt03}.
For the case of \izw, interstellar conditions are quite different from those
in \sbs, in spite of their similar metallicities.
Since the visual extinction $A_V$ is very low even in its star forming 
patches, the radiation reprocessing from UV to FIR is inefficient.
It naturally leads to a low IR luminosity $L_{\rm IR} \la 10^{7.5}\,L_\odot$.
For the same reason, the silicate band features are observed in emission
in the SED of \izw.

It is worthwhile to focus on the differences between \izw\ and \sbs.
Indeed, these two representative low-metallicity star-forming dwarfs are 
significantly different in spite of their almost same metallicities.
First, the SFR of \izw\ is much smaller than that of \sbs. 
This means that the dust production rate of \izw\ is much smaller than 
that of \sbs.
In spite of their optical similarity, \izw\ has rather normal SFR while
\sbs\ has extremely vigorous star formation.
Such an active/passive star formation variety may play an important role
in understanding their activity \citep{hunt02}.
Second, the apparent SF region size of \izw\ is much larger than 
that of \sbs. 
This difference could be produced by the dust sweeping by the radiation 
pressure from the central stellar cluster.
When the age of the system becomes older, the dust layer may be swept
far away from the star-forming complex, and the opacity may get smaller.
Since \izw\ has an older age than \sbs, the dust layer of the \izw\
can be transported further away. 
The small dust opacity of \izw\ might be understood by the combination of 
the above two effects.

Inspection of Figure~\ref{fig:izw} tells us that the peak of the SED
of \izw\ is located around $100\,\mu$m both in the most optimistic and 
pessimistic predictions.
However, the detection in these wavelengths is limited by Galactic cirrus.
A shorter wavelength ($\sim 60\,\mu$m) observation will be dramatically easier.
The MIR observation from space seems exciting, and crucial to examine 
the validity of our prediction. 
In the NIR, dust contribution may be negligible in the NIR.
The fraction of dust emission appears to be smaller than that of \sbs.
This suggests that either of stellar and nebular contribution is dominant
in the NIR.

We discuss the possibility and prospects for future observation in 
the next subsection.

\subsection{Prospects for future observations}\label{subsec:future}

As we have seen in the above, even in the intense UV radiation field of
very young galaxies, small silicate grains suffer from stochastic 
heating and consequently have a broad temperature distribution.
Hence, silicate grains mainly contribute to the MIR continuum emission.
Carbonaceous grains have larger sizes and can be in thermal equilibrium,
but the strong UV field makes their equilibrium temperature very high 
($\sim 50 \mbox{--} 100$~K), and they also contribute to the MIR.
These facts indicate that both (restframe) MIR and FIR surveys are 
necessary to examine the dust amount of galaxies over wide range of 
metallicity.

Our models show that the observed flux of \izw\ is expected to be 
$10\mbox{--}150\,$mJy at $100\,\mu$m. 
This agrees very well with a simple argument by \citet{kamaya01},
who showed that the expected flux for \izw\ is 7--100 mJy at $100\,\mu$m.
Typical detection limits for the forthcoming FIR surveys are
$1\mbox{--}100\,$mJy at $\sim 80\,\mu$m and 
$10\mbox{--}100\,$mJy at $\sim 150\,\mu$m.
We expect $\sim \mbox{several} \times 10^2$ detections of such 
local low metallicity galaxies by some space observations in the near future
such as the {\sl SIRTF}\ SWIRE Legacy program and {\sl ASTRO-F} FIR all-sky survey,
if we assume the local luminosity functions for star-forming dwarf galaxies.
However we should be aware of the cirrus confusion limit, which makes it 
difficult to detect IR extragalactic sources at wavelengths 
$\lambda > 100\,\mu$m.
In order to detect these galaxies at $\lambda = 100\mbox{--}200\,\mu$m, 
a large aperture ($\phi \sim 2\mbox{--}4\,$m) FIR space telescope like
{\sl Herschel} is required to obtain a small diffraction limit to overcome
the cirrus confusion limits.
{\sl Herschel Space Observatory} will be one of the optimal instruments 
for this purpose (on the estimation of the FIR confusion limit for various
forthcoming instruments, see e.g., \citealt{ishii02}).
In any case, such a sample will be useful for the systematic studies 
on `primeval galaxies', through the investigation of other possibly 
`young' local dwarfs \citep[e.g.,][]{hopkins02,corbin02}.

Furthermore, our result for hot dust also offers a potentially important
cosmological insight into high-$z$ galaxies: if dust in high-$z$ object is 
as cool as usually assumed, then we cannot expect a high comoving SFR
density because in a realistic cosmology the cosmic IR background (CIRB) 
spectrum strongly constrains the high-$z$ IR emission \citep{takeuchi01a}.
However, very hot dust in such high-$z$ object reconciles the CIRB 
constraint with high SFR, and allows vigorous star 
formation hidden by dust in the early Universe \citep{totani02}.
Some studies attempt to predict the number of submillimetre galaxies at
high-$z$ ($z=2\mbox{--}3$) from both empirical and theoretical viewpoints
\citep[e.g.,][]{guiderdoni98,tan99,magliocchetti01,takeuchi01b}.
Most models assume galaxy SEDs similar to those of local dusty 
giant starbursts like M82 or Arp 220.
However, we advocate our theoretical framework for very young 
galaxies to make it possible to treat dwarf starbursts, because present
galaxy formation models suggest a vast number of dwarf galaxies in the 
early Universe.
\citet{hiraferrara02} provided a comprehensive framework of 
very high-$z$ (or, say, primeval: $z\ga 5$) dwarfs in the submillimetre 
and millimetre, and showed the ALMA detectability.
They also suggest the existence of
dust warmer than the typical Galactic environment because of
the intense radiation field.
This emphasizes the importance of our SED modeling which is
applicable to young galaxies with intense interstellar radiation fields. 
With our SED modeling, thus, the advent of ALMA will enable us to compare
our models with observations of such low mass, {\sl bona fide} primeval
galaxies.

\section{Conclusions}\label{sec:conclusion}

We have constructed a simple model of mid- to far-infrared (MIR and FIR) 
SEDs of galaxies to investigate the IR properties of very young galaxies,
based on the galaxy evolution model of \citet[][H02]{hirashita02}.

Small grains cannot establish thermal equilibrium with the ambient radiation 
field because of their very small heat capacity.
We find that, even in the intense UV radiation field of very young galaxies,
small silicate grains still cannot be in equilibrium, and consequently have 
a broad temperature distribution.
Hence, silicate grains contribute mainly to the MIR continuum emission.
Carbonaceous grains are larger and can be in thermal equilibrium,
but the strong UV field makes their equilibrium temperature very high 
($\sim 50 \mbox{--} 100$~K); hence they also contribute to the MIR.

We calculated the SEDs of very young, low-metallicity galaxies as a function
of starburst age.
In the first several Myrs, such a galaxy shows very hot dust emission with 
a peak at $\lambda \sim 30\mbox{--}50\,\mu$m, because the self absorption is 
not very pronounced in this phase.
Later, the SED changes its appearance to a more common shape, 
dominated by cooler dust emission, caused by the strong self-absorption of 
MIR light and re-emission at submillimetre wavelengths.
The evolution of the dust opacity and the shape of the IR extinction curve 
were also given.

We succeeded in reproducing the MIR--FIR SED of \sbs, a low-metallicity 
(1/41 $Z_\odot$) galaxy with an unusual IR SED and a strong MIR flux.
\sbs\ is consistent with a very young starburst of age $6.5 \times 10^6$~yr 
and compact size of $\sim 20$~pc.
The model results of young age and compact size are also consistent with 
the optical properties of the galaxy.
We also successfully reproduced the extinction curve of \sbs, which
agrees well with the semi-empirical extinction curve deduced from 
the observational data and our unextincted `bare' continuum SED.
We obtain the visual extinction strength $A_V \sim 18\,$mag, consistently
with Hunt et al.'s estimate of $A_V =16.3\,$mag.

For future observations, we then made a prediction for the SED of 
another extremely low-metallicity galaxy, \izw.
As in the case of the optically visible part of \sbs, \izw\ has a low visual 
extinction, which suggests a small dust opacity.
Hence it follows that the FIR luminosity of \izw\ is probably 
low as $L_{\rm FIR} \sim 10^{7\mbox{--}7.5}\,L_\odot$, 
depending on the observational constraint of dust mass.
We expect the peak of the IR SED of \izw\ to be around $100\,\mu$m, 
but Galactic cirrus may hamper its detection at these wavelengths.
Shorter wavelength observations will be much more effective.

We conclude that both of MIR and FIR surveys are necessary to infer   
the dust amount of galaxies with a wide range of metallicities.
About several hundreds of local low metallicity galaxies may be detected in 
the such as the {\sl SIRTF} SWIRE Legacy program and {\sl ASTRO-F} 
FIR all-sky survey, and even more detailed observations will be possible by 
{\sl Herschel} and other 2--3~m class FIR space telescopes.
Such samples will be useful for the systematic studies on the local
populations of very young galaxies, as well as for their counterparts at 
high-$z$.

In addition, redshifted peaks of very hot dwarf galaxies will be found in the
submillimetre wavelengths ($\sim 50(1+z)\,\mu$m, and $z\ga 5$), and 
we conclude that the advent of ALMA will enable us to observe 
such low mass, {\sl bona fide} primeval galaxies.

\section*{Acknowledgements}
First we thank the anonymous referee, whose careful comments improved 
the quality of this article very much.
We are greatly indebted to Veronique Buat, Denis Burgarella, Alessandro
Boselli, Marc Sauvage, Takashi Onaka, Akio K.\ Inoue, Hideyuki Kamaya, 
Koji S.\ Kawabata, Misato Fukagawa, Yoko Okada, Masato Onodera, 
Motohiro Enoki, and Kohji Yoshikawa for fruitful discussions and comments.
We also acknowledge Trinh X.\ Thuan and Marc Sauvage for kindly providing 
their ISOCAM spectroscopic data of \sbs.
TTT and HH have been supported by the Japan Society of the Promotion of 
Science.

\end{document}